\documentclass[aps,prl,twocolumn,floatfix,letterpaper]{revtex4}
\usepackage{graphicx}
\usepackage{amsmath}
\usepackage{amssymb}
\usepackage{booktabs}
\usepackage{xcolor}

\usepackage{bm}
\usepackage{tabularx}   
\usepackage{MnSymbol,wasysym}
\usepackage{sidecap}

\newcommand{\be}{\begin{eqnarray}}
\newcommand{\ee}{\end{eqnarray}}
\newcommand{\bse}{\begin{subequations}}
\newcommand{\ese}{\end{subequations}}
\newcommand{\bea}{\begin{align}}
\newcommand{\eea}{\end{align}}
\usepackage{graphicx}   

\definecolor{bronzerefs}{rgb}{.53,.33,.04}

\begin{document}

\title{
Self-transport of swimming bacteria is impaired by porous microstructure
}

\author{Amin Dehkharghani$^{1}$, Nicolas Waisbord$^{1,2}$ and Jeffrey S. Guasto$^{1,\ast}$} 
\affiliation{ 
$^1$Department of Mechanical Engineering, Tufts University,
200 College Avenue, Medford, MA 02155, USA\\
$^2$Univ Rennes, CNRS, Geosciences Rennes, UMR 6118,
F35000, France\\
$^\ast$Corresponding author
} 



\begin{abstract}
Motility is a fundamental survival strategy of bacteria to navigate porous environments.
Swimming cells thrive in quiescent wetlands and sediments at the bottom of the marine water column, where they mediate many essential biogeochemical processes.
While swimming motility in bulk fluid is now well established, a comprehensive understanding of the mechanisms regulating self-transport in the confined interstices of porous media is lacking, and determining the interactions between cells and surfaces of the solid matrix becomes paramount. 
Here, we precisely track the movement of bacteria (\emph{Magnetococcus marinus}) through a series of microfluidic porous media with broadly varying geometries and show that cell motility results in a succession of scattering events from the porous microstructure.
Order or disorder can impact the cells' motility over short ranges, but we directly demonstrate that their large-scale transport properties are regulated by the cutoff of their persistent swimming, which is dictated primarily by the porosity and scale of the porous geometry.
The effective mean free path is established as the key geometrical parameter controlling transport, and along with with minimal knowledge of cell swimming motility and surface scattering properties, we implement a theoretical model that universally predicts the effective diffusion of cells for the geometries studied here. 
These results are an important step toward predicting the physical ecology of swimming cells in quiescent porous media and understanding their role in environmental and health hazards in stagnant water.

\end{abstract}

\maketitle

\section{Introduction}
Bacterial communities are abundant in stagnant waters of wetlands including swamps, marshes, bogs, and fens, and they play a key role in biogeochemical cycles by functioning as primary producers as well as decomposers~\cite{Upadhyay2017,Singh2020}.
More than 80\% of the known bacterial species swim by means of thin, actuated long flagella~\cite{Moens1996,Soutourina2001}, and their resultant self-transport is vital to many of their ecological functions, especially in the stagnant porous sediments in which they live.
In bulk fluid, many bacteria swim in persistent random walk patterns~\cite{Berg1972,Polin2009,Taktikos2013}, where otherwise straight-line swimming is decorrelated by flagellar-induced turns and Brownian rotational motion~\cite{berg1993random}.
In porous micro-environments, random walk motility is further disrupted due to cell reorientations caused by nearby solid surfaces \cite{Lanning2002a,Ford2007,RismaniYazi2018,Bertrand2018,Jakuszeit2018, Zhang2014, Tokarova2021}.
While there has been significant emphasis on understanding the physical origins of cell-surface interactions, a clear consensus of their net effect on the self-transport properties of swimming bacteria in porous media has not been established. 
\par
The presence of solid boundaries are known to have significant effects on the swimming trajectories of nearby motile cells due to both hydrodynamic and steric forces~\cite{Lauga2009b}. 
As a result, some spermatozoa and bacteria swim along surfaces for prolonged times, accumulating near flat substrates ~\cite{Rothschild1963, Berke2008}.
Bacteria have also been shown to become trapped in orbits that follow the convex curvature of near circular obstacles ($\sim 100$~$\mu$m)~\cite{Sipos2015}. 
Flagellar contact with a solid substrate, combined with hydrodynamic interactions, were directly shown to mediate surface scattering of plankton~\cite{Kantsler2014,Contino2015}, as well as the motility and tumbling of bacteria~\cite{Li2009,Molaei2014}.
\par
The natural porous habitats of swimming cells have pores that can range in size from that of a single cell to many times larger \cite{Minagawa2008, Yamamoto1988, Nimmo2004}.
Numerical simulations and theoretical works have illustrated the importance of the swimmer interactions with boundaries on their migration through porous media by modeling single microswimmer dynamics~\cite{Chepizhko2013, Morin2017,Jakuszeit2018,Weber2019,Dhar2020}.
In particular, the nature of their scattering boundary conditions (e.g. sliding versus specular reflection) is predicted to have a large impact on their effective diffusion coefficients~\cite{Jakuszeit2018}.
Recent experiments on bacterial self-transport in 3D packed beds show that small pores comparable to the bacterial body size ($\approx$ 1 - 10 $\mu$m) result in a cage-hopping behavior between pores, likely mediated by flagellar contact \cite{Bhattacharjee2019a}.
Measurements of swimming microalgae in regular microfluidic lattices with larger pores (30~$\mu$m) established that increasing the density of solid obstacles reduces the effective diffusion of the cells~\cite{Brun-Cosme-Bruny2019}.
Experiments and simulations for synthetic active particles have revealed the existence of transient subdiffusive transport in random media with low porosity \cite{Morin2017}, but it is unclear whether these results are applicable to biological microswimmers.

\par
The random walk behavior of swimming cells and active particles has inspired a number of diffusive models for transport in crowded environments \cite{Chepizhko2013, Morin2017,Jakuszeit2018,Weber2019,Dhar2020}, most of which are particular to specific propulsion mechanisms, particle-surface interactions, and porous microstructure \cite{Bechinger2016}.
A pervasive assumption in describing swimming cell transport in porous media is that the persistent swimming motion is truncated approximately to the mean pore size of the medium.
Inspired by classic models for gas diffusion \cite{Pollard1948}, a general framework for swimming cell transport in porous media has been suggested~\cite{Ford2007}.
However, these assumptions and models remain largely untested due to a lack of robust experimental measurements of cell trajectories and transport properties.
\par
Here, we investigate the effect of porous microstructure on the self-transport of swimming bacteria (\textit{Magnetococcus marinus}; Fig. \ref{Fig1} A-F) through comprehensive microfluidic experiments.
High-precision cell tracking follows individual cells through ordered and disordered model porous media having porosities and pore sizes spanning natural cell environments~\cite{Al-Raoush2005, Schnaar2006, Minagawa2008, Yamamoto1988, Nimmo2004}.
Direct measurements of cell motility and scattering angles off the solid surfaces connect the micro-scale cell behaviour to their large-scale transport coefficients.
We show how the persistent random walks of the cells are broken by the porous microstructures, reducing the effective diffusion coefficients across all geometries.
These results inform the development of an effective mean free path as a fundamental length scale of the porous media, which enables the accurate prediction of cell transport coefficients through a simple statistical model.



\section{Materials and methods}

\subsection{Cell culture}

\textit{Magnetococcus marinus} (strain MC-1) were grown in a semi-solid medium as previously described~\cite{Waisbord2016a, Bazylinski2013}.
A stock solution was prepared containing the following chemicals and solutions per liter: NaCl, 16.43~g; MgCl$_2\cdot$6H$_2$O, 3.5~g; Na$_2$SO$_4$, 2.8~g; KCl, 0.5~g; CaCl$_2\cdot$2H$_2$O, 1~g; HEPES, 2~g; NH$_4$Cl, 0.3~g; Wolfe's mineral elixir~\cite{Frankel1997}, 5~mL; 0.2\%~(w/v) aqueous resazurin, 0.6~mL; 0.01~M FeSO$_4$, 3~mL; 0.5~M KHPO$_4$ buffer with pH~6.9, 1.8~mL.
The pH of the premade stock solution was adjusted to 7.0 using a 5\% (w/v) NaOH solution, 1.6~g of noble agar (Difco) was added and mixed, and the stock solution was autoclaved, which fully dissolved the agar.
After autoclaving, the stock solution was allowed to cool to approximately 45$^\circ$C.
The following solutions and chemicals were then added to the stock solution (per liter) to make the final cell growth medium: L-cysteine hydrate, 400~mg; vitamin solution~\cite{Frankel1997}, 0.65~mL; 40\% sodium thiosulfate pentahydrate solution (i.e., 10~mM of energy source), 5.6~mL; and 0.8~M NaHCO$_3$ (autoclaved dry; sterile water added after autoclaving to make the fresh stock solution), 2.8~mL. 

For cell cultures, 10~mL of the medium was dispensed into sterile, 15~mm diameter and 125~mm long screw-capped autoclaved glass test tubes, which do not allow ambient air to diffuse into the medium. 
The glass tubes with the medium were tightly capped and refrigerated over night allowing the agar to semi-solidify.
During this time, the L-cysteine reduced the oxygen in the medium and formed an oxygen gradient relative to the air trapped at the top of the tube. 
The oxic-anoxic interface was evident by the transition from a pink hue to a transparent medium from the top of the glass tube to the bottom. 
100 $\mu$L of cells from a previous culture was pipetted at the oxic-anoxic interface to inoculate the medium in each glass tube.
All cultures were incubated at room temperature, and after approximately three days, a microaerobic band of bacteria formed at the oxic-anoxic interface of the tubes. 
For the microfluidic experiments, the cells were re-suspended (see below for details) in a swimming medium that is similar to the growth medium, except that it does not contain agar, HEPES, resazurin, vitamins, L-cysteine, and NaHCO$_3$. 



\subsection{Microfluidic devices and experiments}

\begin{figure*}[t]
\includegraphics[width=1\textwidth]{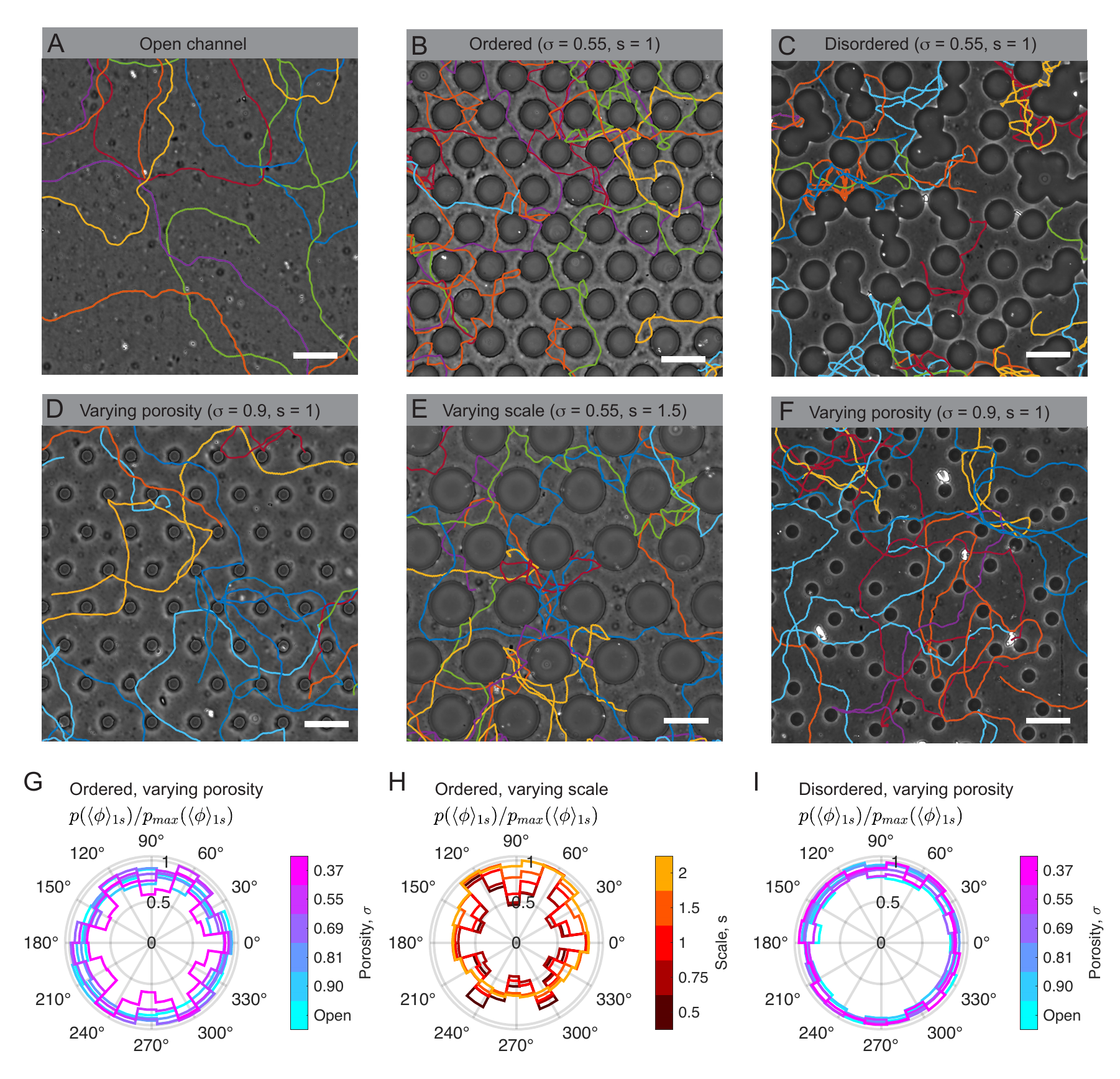}
\caption[Porous microstructure frustrates the random walks of swimming cells.]{Porous microstructure frustrates the random walks of swimming cells. 
(A-F) Sample cell trajectories in porous microfluidic devices for various porosity, scale, and disorder of the obstacle lattice. See electronic supplementary material (Supplementary Table~1) for a full list of porous microfluidic geometries tested. Scale bars, 120~$\mu$m.
(A) Cell trajectories in an open channel ($\sigma =1$).
(B) Ordered porous medium (porosity, $\sigma =0.55$) with cylindrical pillars of diameter $d = 85$~$\mu$m, arranged in a hexagonal lattice with a lattice spacing $c = 120$~$\mu$m (scale, $s = 1$).
(C) Disordered porous medium ($\sigma =0.55$, $s = 1$), where $d$ and average lattice spacing are identical to (B). 
(D) Porosity is varied by modifying the pillar diameters ($d$ = 40 $\mu$m), while holding the lattice spacing constant ($c$ = 120 $\mu$m).
(E) Identical to (B), but with all features scaled up by a factor $s=1.5$.
(F) Identical to (C), but with increased porosity by reducing the pillar diameters to $d = 20$~$\mu$m.
(G - I) Normalized polar probability density of cell swimming directions in various geometries: 
(G) ordered lattice with varying porosity ($s = 1$), (H) ordered lattice with varying scale ($\sigma$ = 0.55), and (I) disordered media with varying porosity ($s$ = 1). 
Components of the instantaneous swimming direction vectors are averaged over a 1~s interval of the cell trajectories to obtain average swimming directions, $\langle \phi \rangle_{1s}$, and their probabilities, $p(\langle \phi \rangle_{1s})$. The latter is normalized by its maximum value, $p_{max}(\langle \phi \rangle_{1s})$.
}
\label{Fig1}
\end{figure*}

Model porous microfluidic devices were fabricated using standard soft-lithography techniques \cite{Whitesides2006}. 
The devices consisted of circular pillars in otherwise rectangular cross section channels with various pillar arrangements (i.e., hexagonal order or random disorder), porosities, and sizes (Fig.~\ref{Fig1} A-F; electronic supplementary material, Table~S1). 
The devices had an inlet and an outlet to fill the chambers with swimming media and bacteria. 
In the ordered porous media, two different approaches were used to vary the size of the lattice constant ($c$) and the pillar diameters ($d$): \textit{i}) varying the porosity ($\sigma$ = 0.90, 0.81, 0.69, 0.55, 0.37) by keeping the lattice spacing constant ($c$ = 120 $\mu$m) and changing the pillar diameters ($d$ = 40, 55, 70, 85, 100~$\mu$m), and \textit{ii}) varying the scale ($s$ = 0.5, 0.75, 1, 1.5, 2) by scaling all the dimensions up or down ($c$ = 60, 90, 120, 180, 240~$\mu$m) at a constant porosity ($\sigma$ = 0.55).
The disordered porous media were generated by randomly displacing the cylindrical pillar locations from an originally hexagonal lattice. 
The new pillar locations were randomly sampled inside a hexagon with a circumradius of 0.9$c$ centered at the original pillar location~\cite{Dehkharghani2019}.
Five different patches of pillar arrangements (2.6~mm long each) were located adjacent to one another inside of a main channel having a total length of 25~mm, width of 2.2~mm, and depth of 10~$\mu$m. 
Open spaces devoid of pillars ($\sigma = 1$) were located near the inlet and outlet of the channel for control experiments.


To perform the experiments, the devices were filled with swimming media, and the outlet of the channel was subsequently sealed with wax to eliminate residual flows.
1~$\mu$L of cell suspension was taken from the oxic-anoxic transition interface of the glass culture tube, and pipetted into the open inlet of the channel. 
North-seeking magnetotactic cells were guided into the channel using a hand held magnet, until the cell concentration reached about 50--100~cells/mm$^2$ by visual inspection. 
The remaining cells at the inlet were removed via pipette to avoid changes in cell concentration during the experiment.

\subsection{Cell imaging and tracking}
Imaging was performed approximately at the center of each of the porous segments within the main channel as well as in the open section of the channel devoid of porous geometry ($\sigma = 1$) and far from the side walls of the devices. 
Cells were imaged using phase-contrast microscopy on an inverted microscope (Nikon Ti-E) with a 4$\times$ magnification objective (0.13 NA).
For each porous geometry, a 3600~frame video was recorded at 45~fps (5MP; Blackfly S camera, FLIR).
This optical arrangement provided a large depth of focus relative to the channel depth and reduced light scattering around cylindrical pillars, and it enabled robust tracking of the $\approx 1$~$\mu$m diameter bacteria throughout the depth of the microfluidic device and during collisions with pillar walls.
Bacteria were tracked using a custom Kalman filter particle tracking algorithm (MATLAB, MathWorks)~\cite{Kalman1960}, yielding $\approx~2,000$ cell trajectories per video.

\subsection{Statistical analysis of cell transport}

The swimming speed of individual cells varies widely among the cell population and is in the range of $20 \le V_s \le 110$~$\mu$m/s (electronic supplementary material, Fig. S1), where $V_s$ is the mean swimming speed of a single cell averaged over its trajectory. 
Variations in the distribution of cell swimming speed significantly affect the perceived transport coefficients due to the strong dependence on $V_s$. 
For example, the effective diffusion coefficient that characterizes the random walk for swimming cells with a persistence time, $\tau_p$, is $D = V_s^2\tau_p/2$.
In order to isolate the effect of swimming speeds from the effects of porous microstructure on cell transport, cell trajectories were separated into three sub-populations throughout our analysis having relatively narrow swimming speed ranges: 20--50~$\mu$m/s, 50--80~$\mu$m/s, and 80--110~$\mu$m/s. 
We primarily focus on the cells with swimming speeds in the range of 50--80~$\mu$m/s, unless stated otherwise.

Mean squared displacements (MSDs) were determined from cell trajectories by dividing the tracks into non-overlapping, equal-time segments equivalent to the time shift, $t$, and computed as $MSD(t) = \langle (x(t + t_0) - x(t_0))^2 + (y(t + t_0) - y(t_0))^2 \rangle$.
The experimental MSDs were fitted with the analytical expression $MSD(t) = 4Dt(1-\exp(-t/2\tau_p))$ to obtain the cells' effective diffusion coefficient, $D$, and persistent swimming time, $\tau_p$~\footnote{This expression is a practical approximation of the full solution of the MSD of a swimmer at constant speed $V_s$, with a decorrelation rate of its orientation $1/\tau_P$: $MSD(t) = 4Dt(1-\frac{1}{t/\tau_P}+\frac{e^{-t/\tau_P}}{t/\tau_P})$}.
The mean swimming speed, $V_s$, is computed by fitting the first 10 points of the MSDs in the ballistic regime ($t \le 0.2$~s), where $MSD = V_s^2 t^2$.
The time correlation function of the cell swimming direction was computed by, $C(t) = \langle \mathbf{p}(t + t_0) \cdot \mathbf{p}(t_0) \rangle$, where $\mathbf{p}$ is a unit vector describing the instantaneous cell swimming direction.
Cell orientation correlation functions decays exponentially, $C(t) = \exp(-t/\tau_p)$, during their random walk motility, which provides an alternative method to compute the persistence time of cell motility and compares well to the MSD fitting approach (electronic supplementary material, Fig.~S2).


\section{Results and discussion}

\subsection{Cell trajectories rapidly decorrelate despite short-time pore-scale guidance}

The long, meandering random walks of swimming bacteria in bulk fluid (Fig.~\ref{Fig1}~A) are drastically altered by interactions with porous microstructure (Fig.~\ref{Fig1}~B-F).
Upon collisions with pillars, cell swimming directions can change abruptly (Fig.~\ref{Fig1}~B-F), which leads to highly tortuous swimming paths in random media (Fig.~\ref{Fig1}~C). 
However, in ordered porous media, cells can briefly follow paths that are aligned with the hexagonal lattice (Fig.~\ref{Fig1}~B).
Beyond the effects of disorder, the observed random walk behaviors of the cells qualitatively appear to be diminished or exacerbated as porosity and feature scale vary (Fig.~\ref{Fig1}~D-F), leading to a vast parameter space whose impact on bacterial transport remains poorly understood.
These fundamental, microstructural parameters were systematically varied (electronic supplementary material, Table~S1) to determine their impact on the short and long time transport properties of swimming bacteria. 
For short times, relative to their bulk persistence time ($\tau_{p,0} = 5$~s), the probability of the swimming direction of the bacteria averaged over 1~s, $p(\langle \phi \rangle_{1s})$, is homogeneous in open space (Fig.~\ref{Fig1} G, I) indicating that the cells swim in random directions (Fig.~\ref{Fig1} A).
However, in ordered media, peaks in the swimming orientation distribution emerge for sufficiently low porosities (Fig.~\ref{Fig1} G), which indicate that the bacteria align with the hexagonal lattice directions.
This short time guiding effect in ordered media is not dictated solely by porosity as it requires sufficiently small scale of the pore features (Fig. \ref{Fig1} H).
While alignment to the local pore structures may persist in random media, the swimming directions remain homogeneous across all porosities examined (Fig.~\ref{Fig1} I) by virtue of the random orientation of pores and throats.

\begin{figure}[t!]
\includegraphics[width=\columnwidth]{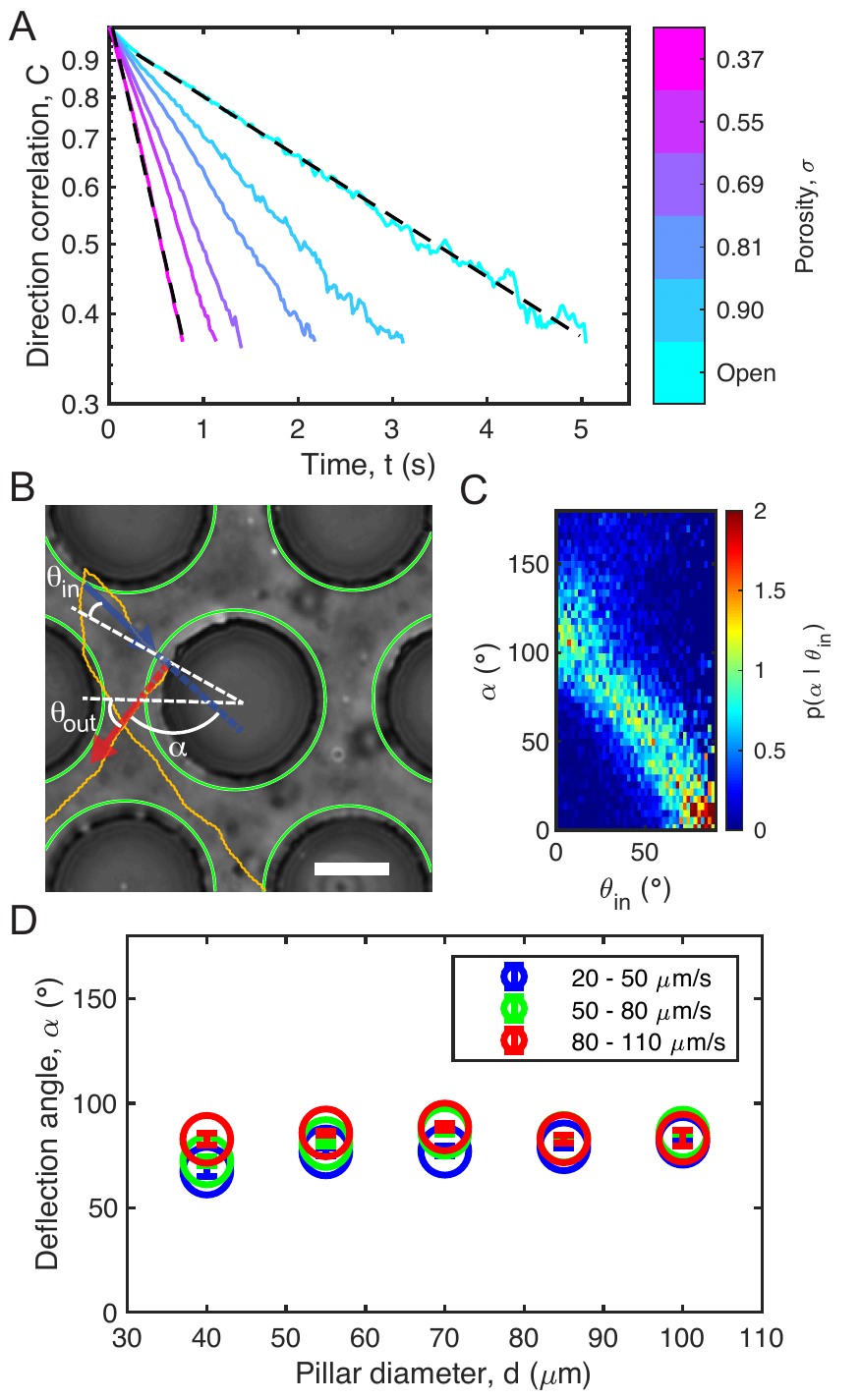}
\caption[Cell swimming direction decorrelates due to scattering from pillar surfaces.]{Cell swimming direction decorrelates due to scattering from pillar surfaces.
(A) Swimming direction correlation functions (see Materials and methods), $C(t)$, for cells in a hexagonal lattice (scale, $s$ = 1) exhibit an exponential decay across a range of porosities ($0.37 \le \sigma \le 1$). Dashed black lines show example fits to $C(t) = \exp(-t/\tau_p)$.
(B)  Yellow dotted line shows a cell trajectory with several scattering events from nearby pillars. The blue vector indicates the measured cell orientation upon entering the impact area (green circle) of the central pillar with an incident angle ($\theta_{in}$) relative to the surface normal (white dashed line). The red vector indicates the cell orientation upon exiting the impact area with an outgoing angle ($\theta_{out}$). The cell orientation changes by a scattering angle, $\alpha$, defined in the range of 0$^\circ$ - 180$^\circ$.
(C) Conditional probability, $p(\alpha | \theta_{in})$, of cell scattering angle given the incident swimming angle measured in a hexagonal lattice ($\sigma$ = 0.37, $s=1$).
(D) Ensemble mean cell scattering angles, $\tilde{\alpha}$, measured directly from for various pillar diameters, $d$, and swimming speed ranges. Error bars show standard error of measured scattering angles. 
}
\label{Fig2}
\end{figure}

Despite the observed sensitivity of the short-time swimming direction to the geometry, cells randomize their direction after a finite time. 
The correlation functions of the swimming direction decay exponentially,  $C(t)=\exp(t/\tau_p)$, for all the inspected geometries (Fig. \ref{Fig2} A), and they are all fully characterized by the persistence time, $\tau_p$.
This exponential decorrelation of the swimming direction stems from both the intrinsically noisy trajectories of the bacteria, which can be observed in the open channels, as well as their encounters with solid surfaces. 
As the porosity is decreased, the rate of such encounters is increased and the swimming direction of the cells decorrelates faster (Fig. \ref{Fig2} A).
Taken together, these observations suggest that the magnitude of the pore size -- rather than porosity or feature size scale alone -- regulate bacterial self transport. 
To fully understand this process, it is necessary to quantify the nature of the decorrelation process due to bacterial interactions with the porous microstructure.


\begin{figure*}[t!]
\includegraphics[width=1\textwidth]{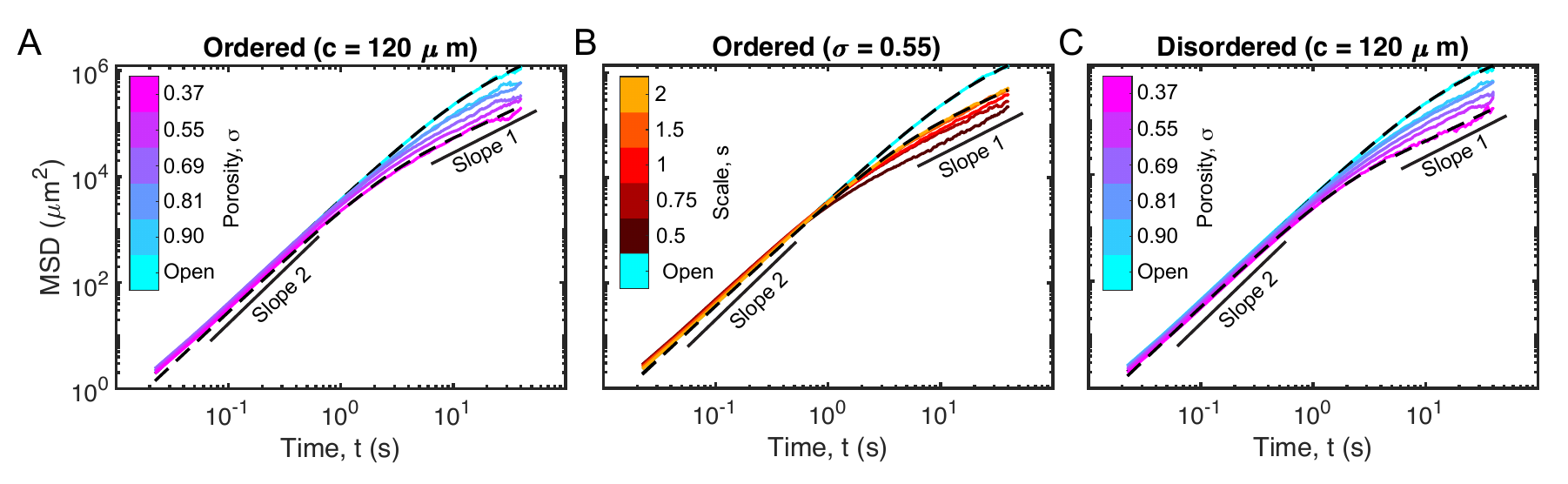}
\caption[Porous media universally hinders cell transport.]{Porous media universally hinders cell transport.
Measured mean squared displacements (MSDs) for swimming bacteria ($50 \le V_s \le 80$~$\mu$m/s) in (A) ordered lattice with varying porosity, (B) ordered lattice with varying scale, and (C) disordered lattice with varying porosity.
In (A-C), the dashed black lines show two examples each of fitted analytical model for the MSD of a persistent random walk (see Materials and methods), and solid line segments indicate the ballistic and diffusive regimes where the MSD grows as $\sim t^2$ and $\sim t$, respectively.
}
\label{Fig3}
\end{figure*}

\subsection{Bacterial scattering is independent of pillar size and cell swimming speed}


Robust cell trajectories from experiments enable us to quantify the bacterial scattering angle upon collision with a pillar surface (Fig.~\ref{Fig2} B), which is paramount to understanding the persistent motility and transport of bacteria in porous media.
Focusing on hexagonal lattices, cell collisions with individual pillars were identified by defining an impact region within 2.5~$\mu$m from the pillar surface (comparable to the cell size).
From the cell trajectories, the incident, $\theta_{in}$, and outgoing, $\theta_{out}$, angles of the cells relative to the pillar surface normal (Fig.~\ref{Fig2} B) were measured~\cite{Contino2015} by
fitting straight lines to the cell position over seven frames just prior to entering and just after exiting the impact region, respectively.
The scattering angle, $\alpha = \pi - (\theta_{out} - \theta_{in})$, characterizes the change in cell orientation due to a collision with a pillar (Fig.~\ref{Fig2} B) and empirically appears to depend on their incident angles, $\theta_{in}$, prior to impact (Fig.~\ref{Fig2} C).
Our results show that, when a cell approaches a pillar surface at a glancing, tangent angle ($\theta_{in} \approx 90^\circ$), the cell's orientation generally exhibits little change, and it leaves the surface oriented in approximately the same direction compared to its incident angle ($\alpha \approx 0^\circ$; Fig.~\ref{Fig2} C). 
However, when a bacterium approaches perpendicular to the pillar surface ($\theta_{in} \approx 0^\circ$), it is strongly scattered with a significant reorientation of the cell trajectory ($\alpha \approx 120^\circ$).
The scattering angle decreases approximately linearly with incident angle (Fig.~\ref{Fig2} C). Importantly, we note that these scattering properties likely vary across microbial species due to flagellation, swimming style, and body shape and size~\cite{Contino2015}.
The cell scattering measurements were performed \textit{in situ} across all pillar diameters tested (electronic supplementary material, Table~S1) and three swimming speed ranges (electronic supplementary material, Fig.~S1).
Ultimately, the ensemble average scattering angle, $\tilde{\alpha}$, for the bacteria studied here was found to be independent of both the pillar diameter and the cell swimming speed (Fig.~\ref{Fig2} D).
These measurements not only illustrate the nature of the primary decorrelation mechanism of cell motility in porous media, but also provide key quantitative information necessary to construct a comprehensive model of cell transport in such microstructured environments.

\subsection{Porous microstructure hinders cell transport}

In order to quantify the effects of porous media on the transport properties of swimming bacteria, we analyze the mean squared displacements of the cells across a wide range of pore structures.
For short times ($t \ll \tau_p$), the MSDs universally overlap and grow as $\sim t^2$ due to ballistic cell swimming motility, prior to scattering from pillars, or otherwise reorienting due to flagellar noise or rotational Brownian diffusion (Fig.~\ref{Fig3}).
The scaling of the MSDs eventually transition to $\sim t$ at large times ($t \gg \tau_p$) for all geometries, which is indicative of diffusive transport and stems from the decorrelation of the cell swimming trajectories (Fig.~\ref{Fig2} A).
The MSDs for the 2D swimming trajectories within the imaging plane of the microfluidic devices are well captured by the expression (Fig. \ref{Fig3})~\cite{Patteson2015, Bechinger2016}, $MSD = 4Dt \left[1-\exp(-t/2\tau_p) \right]$, where a fit of the model (Fig.~\ref{Fig4} A--C, dashed black curves) provides the effective diffusion coefficient of the cells, $D$, as well as the persistent swimming time, $\tau_p$. 
The MSD transition time ($= 2 \tau_p$) -- and thus the persistence time -- decreases with both decreasing porosity and scale for all of the pore geometries and swimming speed ranges examined, and provides an alternative manner to realize the reduction in the persistent swimming time (Fig.~\ref{Fig2} A) due to scattering events.
The persistence times measured using this method closely correspond to those measured from the exponential decay of the swimming direction correlation functions, confirming the robustness of the analysis (electronic supplementary material, Fig.~S2).
While the MSDs of the swimming cells coincide for short times, a decrease in the porosity and/or scale causes a more rapid decorrelation of cell motility and a decreased growth of the MSDs for $t\gg \tau_p$. 
This effect is universal across all geometries and bacterial swimming speed ranges (Fig.~\ref{Fig3}) and indicates that the porous microstructure hinders the diffusive transport of the cells, even in the ordered hexagonal geometries that exhibit short-time preferential swimming directions (Fig.~\ref{Fig1} G--H).

\subsection{Effective pore size sets the effective cell diffusion}

\begin{figure*}[t!]
\includegraphics[width=1\textwidth]{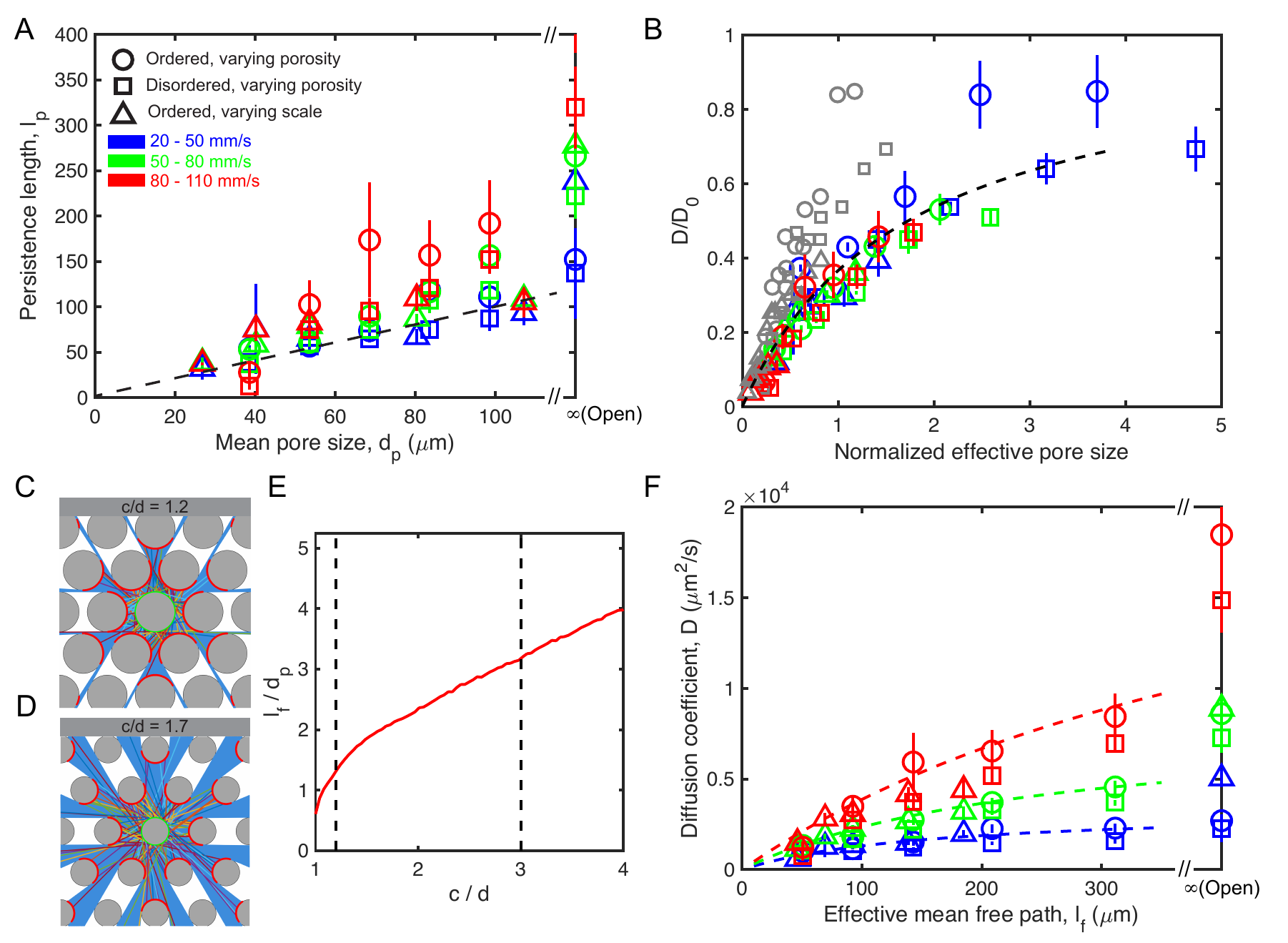}
\caption[Pore geometry sets persistent length and regulates diffusive self transport of swimming cells.]{Pore geometry sets persistent length and regulates diffusive self transport of swimming cells. 
(A)
Persistent swimming length of the cells, $l_p$, decreases with pore size, $d_p$.
Dashed line shows $l_p = d_p$ for reference.
(B)
Normalized effective cell diffusion coefficient, $D/D_0$, decays with normalized effective pore size. 
$x$-axis corresponds to normalized mean pore size, $d_p V_s/D_0$, and normalized mean effective path, $l_f V_s/D_0$, for grey and colored markers, respectively.
$D_0$ is the bulk diffusion coefficient, and dashed line is  Eq.~\ref{eq:MTBNoField_normalizedBosanquet}.
(C - D) 
Effective mean free path is defined as the mean distance for randomly sampled straight paths (colored lines) from the central pillar (green) to the first intersecting pillar (red) in a hexagonal lattice.
Samples are shown for pillar spacing to diameter ratios of $c/d = 1.2$ in (C) and $c/d = 1.7$ in (D). Blue areas indicate coverage of all possible free paths.
(E)
Effective mean free path, $l_f$, in the hexagonal lattice is systematically larger than the mean pore size, $d_p$, within the range of pillar spacing to diameter ratios, $c/d$, explored in microfluidic experiments (dashed lines).
(F)
Effective cell diffusion coefficients, $D$, decrease with decreasing effective mean free path, $l_f$, for all swimming speed populations and geometries. 
Dashed lines show Eq.~\ref{eq:MTBNoField_normalizedBosanquet}.
In (A), (B), and (F), markers correspond to key in (A), and error bars are the standard error computed by dividing the experimental data sets into random subsets ($N=3$).
}
\label{Fig4}
\end{figure*}

The effective cell diffusion coefficients decrease markedly by more than 500\% across the porous geometries investigated (Fig.~\ref{Fig4} B,F).
For their persistent random walk in 2D, the long-time diffusion coefficient of the cells is $D =V_s^2\tau_p/2$, which can alternatively be expressed in terms their persistent swimming length $l_p = V_s \tau_p$ as $D = V_s l_p/2$.
Cell populations with higher swimming speeds exhibit higher diffusion coefficients across all of the geometries examined (Fig.~\ref{Fig4}~F).
Swimming speed, $V_s$, is an intrinsic property of the cells and does not vary with changes in geometry. 
However, the bulk persistent swimming length, $l_{p,0} = V_s \tau_{p,0}$, sets a length scale below which we expect the porous media to dictate $l_p$ (Fig.~\ref{Fig4} A; electronic supplementary material, Fig.~S3).
Our experiments clearly illustrate that ordered versus disordered porous media only marginally affect bacterial transport coefficients (Fig.~\ref{Fig4} A--B, F).
Therefore, the effective pore size is the key parameter in setting the persistent swimming length (Fig. \ref{Fig4} A) and transport coefficients (Fig. \ref{Fig4} B) of the cells and not the porosity, scale, or order of the media.

Apart from random reorientations due to Brownian rotation and self-propulsion noise, motile bacteria in porous media generally swim in straight trajectories prior to interacting with a solid surface.
The mean pore size, $d_p$, is a common geometric measure of the typical distance between such surfaces, and it is often quantified by the mean of the largest circle diameter (or sphere in 3D) that can be inscribed in the void space~\cite{Bhattacharya2006, Roozbahani2017,Chiang2009,Nishiyama2017}.
For example, the mean pore size for a hexagonal lattice is $d_p = 2c/\sqrt{3} - d$, and does not vary significantly for the disordered media studied here.
As the mean pore size decreases, reorientations due to the cell-surface interactions becomes more prominent and decrease the measured persistent swimming length (Fig. \ref{Fig4} A) and time (electronic supplementary material, Fig. S3).
While the persistence length might be expected to be governed by $l_p = d_p$ (Fig.~\ref{Fig4} A, dashed line), cells with higher swimming speeds exhibit persistence lengths that greatly exceed the pore size, especially for larger $d_p$.
In the case of highly porous materials with large $d_p$, the probability of interacting with a small obstacle, $d$, compared to a relatively large obstacle separation, $c$, diminishes, allowing cells to transit multiple pores before scattering.
Thus, we introduce an alternative characteristic length scale, stemming solely from the porous media geometry: 
The effective mean free path, $l_f$, is the average line-of-sight distance from randomly sampled straight paths between circular obstacles (Fig. \ref{Fig4} C-D), and it is akin to trajectories sampled by swimming cells with high persistence.
The effective mean free path relative to mean pore size increases monotonically with $c/d$  and generally $l_f/d_p > 1$ for the diverse media explored here (Fig.~\ref{Fig4}~E).
Below, we show that this intuitive quantity is adept at describing and predicting the transport of swimming cells in porous media (Fig.~\ref{Fig4}~B and F).

\subsection{Predictive model for bacterial transport in porous media}


Based on the classic models for gas diffusion \cite{Pollard1948}, a general statistical framework for modeling the decorrelation cell swimming motility porous media has been suggested~\cite{Ford2007}, but not rigorously explored experimentally.
Here, we reinterpret this model through the lens of our extensive experimental measurements and establish the appropriate geometric length scale that regulates the decorrelation of bacterial motility, and thus their transport, in porous media. 
The total decorrelation rate of the bacterial swimming direction in porous media arises from the sum of two independent processes:
\begin{align}
    \frac{1}{\tau_p} = \frac{1}{\tau_{p,0}} + \frac{1}{\tau_{m}}.
    \label{eq:MTBNoField_Bosanquet}
\end{align}
The first process is the natural decorrelation rate, $1/\tau_{p,0}$, due to rotational Brownian diffusion and flagellar noise, which is related to the effective diffusion coefficient and persistence length in the bulk by $D_0 = V_s^2\tau_{p,0}/2 = V_s l_{p,0}/2$, and  measured directly from experiments (Fig. \ref{Fig4} F).
The second process stems from bacterial scattering events due to interactions with solid surfaces
\begin{equation}
    \frac{1}{\tau_{m}} = \frac{(1-\langle\cos(\alpha)\rangle)}{l_m/V_s},
\end{equation}
It is mathematically similar to a run-and-tumble random walk \cite{berg1993random}, and depends upon the cell swimming speed, $V_s$, and surface scattering angle, $ \langle\cos(\alpha)\rangle $. 
Both of these parameters are measured directly from experiments, where we also show that the scattering statistics are independent of obstacle size (Fig.~\ref{Fig2}). 
Importantly, this decorrelation rate is also a strong function of the effective pore size of the media, $l_m$, which dictates the typical distance that a cell can swim unimpeded.
The resulting effective diffusion coefficient, $D$, for an active swimmer can thus be expressed as
\begin{align}
    D(l_m) = \frac{D_0}{\left(1 + \frac{l_{p,0}}{l_m}(1-\langle\cos(\alpha))\right)},
    \label{eq:MTBNoField_normalizedBosanquet}
\end{align}
which is shown in normalized form in Fig.~\ref{Fig4} B (dashed line).
\par
The mean pore size is a typical choice for the relevant physical length scale of the media, $l_m = d_p$: Indeed it yields a partial collapse of the data (Fig.~\ref{Fig4} B, grey markers) and is captured by the model (Eq.~\ref{eq:MTBNoField_Bosanquet}) at small pore sizes, where cell-surface interactions are frequent.
However, this local measure fails to predict effective diffusivity for large pore size.
In highly porous environments, motile cells can travel several pores prior to interacting with small or sparse obstacles.
Rather, the effective mean free path, $l_f$, computed purely from the geometry of the medium (Fig.~\ref{Fig4}~C-E), provides a more relevant physical length scale, especially for swimmers with high bulk persistence ($l_{p,0} \ge d_p$).  
Indeed, choosing the effective pore size as $l_m = l_f$ shows excellent quantitative agreement of the model with experiments across the broad ranges of porosities, pore sizes, order/disorder, and cell swimming speed ranges examined (Fig.~\ref{Fig4} B, colored markers; Fig.~\ref{Fig4} F).
Thus, these results establish a comprehensive framework for predicting the transport coefficients of swimming cells in porous media with minimal parameters, including: (\emph{i}) the cell motility in bulk fluid, $V_s$ and $\tau_{p,0}$, (\emph{ii}) the average cell deflection from solid surfaces, $\langle\cos(\alpha)\rangle$, and (\emph{iii}) the effective mean free path of the porous medium geometry, $l_f$.

\section{Summary and conclusion}
Our comprehensive experiments illustrate how porous microstructure universally hinders the effective diffusion of swimming bacteria (MC-1), and they establish the essential parameters required to model and predict cell transport in stagnant porous media.
Varying the geometry, including porosity, scale, and randomness, revealed that the reduction of persistence length is the key mechanism regulating hindered transport, and the effective mean free path of porous medium (Fig. \ref{Fig4} C - E) is the primary geometric scale that dictates the bacterial transport coefficients. 
Precise experimental measurements of cell swimming behavior in the vicinity of surfaces were crucial to understanding their long-time transport:
Analysis of  MC-1 cell scattering angles off the pillars indicated that the mean deflection angle was independent of the pillar diameter.
Subsequent scattering events reduced the effective cell diffusion coefficients across a broad range of swimming speeds: 
The fastest cells, which have the largest bulk persistence swimming length, experienced the most significant reduction, a prediction inspired by a modified Bosanquet model \cite{Ford2007}.
The agreement of this model with the extensive direct measurements of diffusion coefficients throughout our experiments illustrates a reliable approach to forecast cell transport properties through porous media.
Without any fitting parameters, this model only requires knowledge of the cell swimming speeds, bulk persistence time, and scattering signature from surfaces along with the effective mean free path of the porous medium, each of which are easily measured or calculated.

Microbes display a host of different behaviors mediated by steric, hydrodynamic, and flagellar interactions with surfaces~\cite{Berke2008}, including a surprising bouncing-like motion in the case of MC-1~\cite{Zhang2014}.
A natural extension of the present work will be to determine how the transport of diverse microbial species with unique swimming strategies and surface behaviors~\cite{Tokarova2021} are affected by porous microstructure through the framework established here.
The effect of directed motility on cell transport through porous microstructures is highly relevant to microbial ecology in sediments and soils, and the approaches described here may provide new insights into magnetotaxis~\cite{Waisbord2021} and chemotaxis~\cite{DeAnna2020} in porous media.
Enhanced transport via preferential cell swimming directions in ordered media have been reported in previously~\cite{Liu1995,Jakuszeit2018}. 
While not observed in the present work, other ordered arrangements and pillar shapes can be investigated to determine the precise conditions under which preferential swimming paths form to amplify transport.
Although the present experiments and modeling were restricted to quasi-2D microfluidic geometries, recently developed experimental approaches~\cite{Bhattacharjee2019} could extend this work to model the naturally occuring 3D environments of swimming microbes.

\subsection*{Acknowledgements}
We thank Thomas Coons for assistance with preliminary experiments that helped to shape the work presented here. 
This work was supported by National Science Foundation awards (to J.S.G.) CBET-1511340, CAREER-1554095, CBET-1701392, OCE-1829827, and CMMI-2027410, as well as MicroMix MSCA-IF-898575 (to N.W.).
\subsection*{Author contributions}
A.D., N.W., and J.S.G. designed the research, developed theoretical analysis, and wrote the paper.
A.D. performed the experiments.
A.D. and N.W. analyzed the experimental data. 

\subsection*{Data accessibility}
Data and materials are available upon request from the authors.

\subsection*{Additional information}
Correspondence and requests for materials should be addressed to Jeffrey.Guasto@tufts.edu.

\subsection*{Competing financial interests} 
The authors declare no competing financial interests.

\bibliographystyle{unsrtnat}
\bibliography{References.bib}

\clearpage

\end{document}